\documentclass[pre,a4paper,twocolumn,footinbib,showpacs]{revtex4}

\usepackage{amssymb}
\usepackage{amsmath}
\usepackage{amsbsy}
\usepackage{graphicx}

\begin{document}
\title{Stretching an adsorbed polymer globule}

\author{Franck Celestini}
  \affiliation{Laboratoire de Physique de la Mati\`ere Condens\'ee, UMR 6622, CNRS, Universit\'e de Nice Sophia-Antipolis,
   Parc Valrose, 06108 Nice Cedex 2, France}

\author{Thomas Frisch}
   \affiliation{
   Institut de Recherche sur les Ph\'enom\`enes Hors \'Equilibre,
   UMR 6594, CNRS, Universit\'e d'Aix-Marseille, Marseille, France}

\author{Xabier Oyharcabal}
   \affiliation{
   Institut de Recherche sur les Ph\'enom\`enes Hors \'Equilibre,
   UMR 6594, CNRS, Universit\'e d'Aix-Marseille, Marseille, France}

\date{\today}

\begin{abstract}
Using molecular dynamic simulation, we study the stretching  of an
adsorbed homopolymer in a poor solvent  with one end held at a
distance $z_e$ from the substrate. We measure the vertical force
$f$ on the end of the chain as a function  of the extension $z_e$
and the substrate  interaction energy $w$. The force reaches a
plateau value at  large extensions. In the strong adsorption
limit, we show that the plateau value increase linearly in $w$ in
good agreement with a theoretical model. In the weak adsorption
limit, a polymer globule with a layered  structure is formed and
elastically  deformed  when stretched. In both cases a simple
theoretical model permits us to predict the relation between
 the necessary force to fully detach the polymer
 and its critical extension.

\end{abstract}


\maketitle
 With the new development  of single molecule experiments, the
 nano-manipulation of individual polymer chains and biological macromolecules  is becoming a
 very important subject
in order to understand their mechanical  properties and
characterize  the intermolecular interactions. Forces on the scale
of the pico-newton have been measured with imposed deformation on
the scale of the nanometer with atomic force microscope (AFM)
\cite{rief99} or optical tweezers \cite{smith01}. Experiments with
AFM have been conducted with different kinds of  biological
molecules such as DNA \cite{strick00}, Titin \cite{rief97} and
more recently Myosin \cite{schwaiger02}. Although experiments on a
flexible polymer are more difficult to conduct  due to its small
persistence length, there have been several recent studies in both
good and poor solvent conditions \cite{chatellier98b,haupt02}.
They show that a stretching experiment can give insight on the
determination of different mechanisms of adsorption and on the
intermolecular forces inside the macromolecule. On the theoretical
side, the stretching of chains in a poor solvent has been studied
by Halperin {\it {et al}} \cite{halperin91b} who predicted that at
weak extensions the spherical globule  deforms into an ellipse and
that for a finite stretching force there exists a sharp first
order  unwinding transition between a globule and a chain.
Numerical and theoretical studies have confirmed this scenario and
have shown that there exists a co-existence state between a
globule and a stretched chain
\cite{wittkop96,lai96,maurice99,kreitmeier99,frisch02a,grassberger02,marenduzzo03,cooke03}.
As shown in \cite{grassberger02} the critical value of the
 force  saturates toward a finite value when the polymerization index $N$ goes to
infinity. In the presence of an adsorbing substrate
force-extension relations have been obtained theoretically
 for an ideal gaussian chain \cite{klushkin02}, for a poly-electrolyte in good solvent
\cite{chatellier98a} and for a polymer chain in a good solvent
\cite{haupt99}. More recently, experimental results have been
obtained for a stretched polymer chain in a poor solvent adsorbed
on a substrate \cite{haupt02}.
\begin{figure}
    \includegraphics[width=0.25\textwidth]{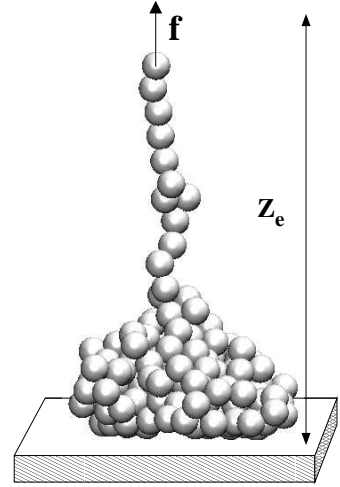}
    \caption{Snapshop  of the polymer  adsorbed on the surface with
    a fixed end (Simulation at $w=4$, $T=0.8$, $N=100$ and $z_e=15$).}
    \label{fig1}
\end{figure}
In this paper, we consider the problem  of an adsorbed polymer
globule in a poor solvent  with one end fixed at a distance $z_e$
perpendicular to the  adsorbing substrate. The substrate is
characterized by an homogeneous interaction energy $w$ with the
polymer. We mainly investigate the effet of strong adhesion $w$,
however we have also perform simulation for intermediate
adsorption for which the onset of poor solvent effect are present.
We do not investigate  here the weak adsorption regime for which
most of the chain can evaporate from the substrate. Simulation
results for the retracting force on the fixed end as a function of
$z_e$ and $w$ are compared with an analytical model. As we stretch
the polymer, we observe that the force rises until it reaches a
plateau value which depends on $w$. 

We use a molecular dynamics simulation technique where the
equations of motion are integrated with the Verlet algorithm and
the temperature fixed with a Nos\'e-Hoover thermostat
\cite{frenkel96}. The interactions between monomers consist of two
parts : $V(r) = V_1(r)+V_{2}(r)$ ($r$ is the distance between two
monomers). $V_1(r)=a(r-d_0)^2+b(r-d_0)^4$ is the valence
interaction between nearest-neighbors (NN) monomers and
$V_{2}(r)=4\epsilon((\frac{\sigma}{r})^{12}
-(\frac{\sigma}{r})^{6})$ the Lennard-Jones interaction between
non NN monomers which englobes the poor solvent effect. Here for
simplicity surface tension effects do not depend on  the polymer
configuration, although this dependence can be taken into account
as done in \cite{frisch02b}, using many body forces. Finally, we
simulate the presence of the atomically flat substrate by adding
an attractive interaction between the wall and the monomers :
$ V_w(z)=
 w\left((\frac{\sigma_a}{z})^9-(\frac{\sigma_a}{z})^3\right)
 $, where
 $z$ is the distance to the wall.  In the following we take
$\epsilon=\sigma=d_0=1$, $a=3000$ and $b=10000$. The energies and
distances are respectively expressed in units of $\epsilon$ and
$\sigma$ and the time step is $\Delta t=10^{-3}$. For the present
study we take $k_b T=0.8$ so that the simulated polymer is under
poor solvent conditions and still above the freezing temperature
\cite{frisch02b}.
 We simulate a polymer adsorbed on the substrate with one end fixed at a distance
$z_e$ from the wall (Fig. 1). In a first run, starting from a free
adsorbed globule, a constant force (usually $f$=10) is applied to
the end of the chain until it reaches the value $z_e$. A second
run, of roughly $10^6$ molecular dynamics (MD) steps with $z_e$
fixed, ensures thermal equilibrium. Finally, during a third run of
$10^7$ MD steps, different quantities like the time-averaged force
on the end of the chain $f$, the density and energy profiles are
recorded. As shown on Figure 1, the  adsorbed globule has  a
semi-spherical shape
due to the effect of the poor solvent.\\
\begin{figure}
 \includegraphics[width=0.37\textwidth]{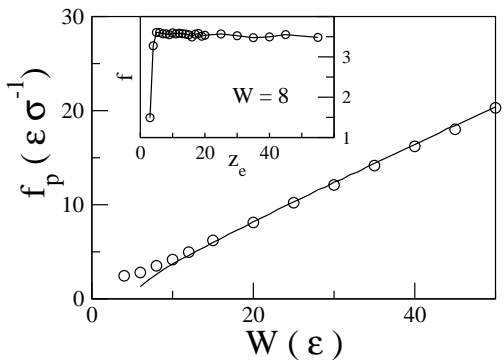}
\caption{Graph of $f_p$ as a function of $w$ for $T=0.8$. Circles
are simulation results of the plateau value. The line corresponds
to the theoretical prediction obtained by minimizing $F$ with
respect to $N_c$. The inset is a force-extension diagram for
$w=8$.} \label{fig2}
\end{figure}
We consider a polymer composed of $N=100$ monomers for which we
perform a set of force-extension measurements for a broad range of
values  of $w$ $(2<w<80)$ and  for $z_e$ varying between $3$ and
$z_{e}^{c}$, the critical value at which the polymer fully
detaches from the substrate. For all $w$, we observe that the
force rises as we increase $z_e$ and saturates to a force plateau
value $f_p$ for the large $z_e$ values. This is illustrated in the
inset of Figure 2 where a full force-extension curve is plotted.
It is then possible to plot $f_p$ as a function of $w$ (Fig. 2).
We observe that in the strong adsorption regime, the dependance of
the plateau in $w$ is linear.  For weaker $w$ values, the poor
solvent effects become prominent and the relation between $f_p$
and $w$ deviates from linearity. \\In Figure 3, we present a
force-extension diagram in the regime of strong adsorption
($w=60$) that presents pronounced oscillations. For these large
values of $w$, the un-stretched part of the polymer is fully
squashed on the substrate and forms a bi-dimensional (2D) globule.
These oscillations are due to the discrete character of the chain
and the peaks correspond to the transfer of a monomer from the
globule to the chain. As we increase $z_e$, $f$ increases rapidly
until the next monomer can escape from the 2D globule. The
stretching tension is then released, and $f$ diminishes abruptly.
As we still increase $z_e$ this process repeats as long as there
are enough monomers adsorbed. $N_c$ being the number of monomers
in the chain, we furthermore observe a dynamical coexistence
between states with $N_c$ and $N_{c}+1$ monomers in the chain.
This is illustrated in the inset of Fig. 3 where we plot the
dynamical variation of $f$ together with its associated
probability density function. For a fixed value of $z_e=18$, the
system can be in two different states with $N_c=18$ and $N_c=19$.
As a consequence the distribution function of $f$ is bimodal.
 For smaller values of $w$ the
morphology of the adsorbed polymer is qualitatively different. As
shown in Figure 1, in the weak adsorption regime the polymer forms
a 3D globule partially wetting the substrate. As we will  discuss
below, the top of the globule which is in contact with the chain
is elastically deformed. As a consequence, the necessary force to
extract a monomer varies continuously and the oscillations in the
force-extension curve almost disappear (inset of Fig. 2).

\begin{figure}
    \includegraphics[width=0.35\textwidth]{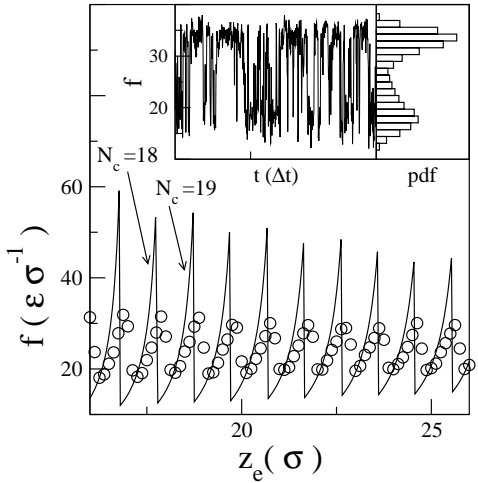}
    \caption{Force-extension diagram in the strong
    adsorption limit for $T=0.8$ and $w=60$. Circles are  simulation results and the line is the
    theoretical  prediction presented below. Inset: $z_e=18$,
   time variation of the force together with its
    associated probability density function illustrating the dynamical co-existence
   between the two states with  $N_c=18$ and $N_c=19$ monomers.} \label{fig3}
\end{figure}

We present a phenomenological model in order to
understand the different
 results reported above. We assume
 that the fixed force and fixed $z$ ensembles are
 equivalent. This is in  principles valid, only in the
 thermodynamic limit. We therefore assume that the system is big
 enough to have small fluctuations.
As shown in Fig 1, the  globule-chain system can be described as a
stretched chain in co-existence with an  adsorbed globule. There
are two main contributions to the  free energy $F$ of the system :
the free energy of the stretched chain $F_c$ and  the free energy
of the adsorbed globule $F_g$. Let $N_g$  and $N_c$ be
respectively the number of monomers in the adsorbed globule and in
the chain, so that $N=N_c+N_g$.  The major contribution to $F_c$
comes from the reduction of the fluctuations of the stretched
chain \cite{flory69}
\begin{equation}
F_c= -N_c k_b T \log{ \left(4\pi \frac{k_b T}{f d_0} \sinh{(f
d_0/k_b T)}\right)} +f z_e \, ,
\end{equation}
where the potential energy of the chain is neglected. The dominant
part of  $F_g$ comes from the interaction energy of the monomers
with the substrate and the interaction between the monomers. We
can neglect here the entropic contribution to  $F_g$
\cite{grosberg96}.
 In the regime of strong adsorption, the interaction between
monomers within the 2D squashed globule can be neglected relative
to the substrate interaction. For large $w$ values we have : $F_g
= - N_g w_{min}$ where $w_{min}$ is the substrate potential
minimum ( $w_{min} \simeq 0.38 w$).
 We write the total free energy $F=F_c + F_g$ where $z_e$
is fixed and $f$ is an implicit function of $z_e$ and $N_c$. Here
$z_e$ is given by the non-linear Langevin relation \cite{flory69}:
\begin{equation}
z_e = N_c d_0 [\coth(\frac{fd_0}{k_bT})-\frac{k_bT}{fd_0}]
\label{langevin1} \, .
\end{equation}
We express the free energy $F$ as a function of the unknown $N_c$
by  inverting numerically  equation (\ref{langevin1}) for $f$. We
then  minimize  numerically the free energy $F$ with respect to
$N_c$ by keeping $w$ and $z_e$ fixed. For a large value of $z_e$,
we substitute in equation (\ref{langevin1}) the obtained value of
$N_c$ to get the plateau force $f_p$. The simulated $f_p$ values
presented in Figure 2 are in good agreement with our theoretical
model for large $w$.  In the high $w$ limit and for $T=0$, we
expect $f_p$ to scale linearly with $w$ so that $z_e=N_c d_0$. In
this case the entropic contribution can be neglected relative to
the  free energy of the globule energy so that $f_p \propto 0.38
w$ in relative good agreement with the simulations ($f_p \propto
0.40 w$ for $T=0.8$). In the weak adsorption limit, the model
underestimates $f_p$ since the interaction between the monomers
has not been included in $F_g$. In this case one should add  to
$F_g$ monomer-monomer interactions of the wetting globule together
with surface tension corrections.   The surface tension correction
gives a  contribution of the order of $\gamma (N-N_c)^{2/3}$ to
$F_g$, where $\gamma$ is a phenomenological surface tension
parameters. This contribution plays only a minor role. We find by
performing the same minimization  as above that there is a very
slight decrease of the plateau value as $z_e$ increased, this is
also confirmed by our MD simulation. A good agreement between
simulation and theory is also found for the extension-curve of
Fig. 3 obtained for various values of $z_e$ using the same
procedure as above. The smoothing out of the peaks is due to the
dynamical coexistence between states with $N_c$ and $N_c+1$
monomers in the chain. This dynamical coexistence  is induced by
the  thermal fluctuations and gives rise to important time
variations of $f$.\\
As we increase the value of $z_e$, we assume that the polymer fully detaches from
the substrate when
 $N_c=N$. Using the relation between $f$, $z_e$ and $N_c$ given in
equation (\ref{langevin1}), we obtain the relation between the
force $f$ and the critical extension $z_e^c$ at which the polymer
detaches from the substrate. We plot in Fig. 4 the theoretical
curve $f(z_e^c)$ and the  simulation results for different values
of $w$ and different system sizes. Here again a good agreement is
found between theory and simulation for the large $w$ values. The
agreement is  slightly less satisfactory for weaker $w$ values. In
this case, we observe that the detachment occurs  suddenly when a
finite number of monomers are still adsorbed on the
substrate($N_c<N$). This is analogous to the first-order unwinding
transition observed on a globule under tension and discussed in
the introduction.
\\
\begin{figure}
    \includegraphics[width=0.4\textwidth]{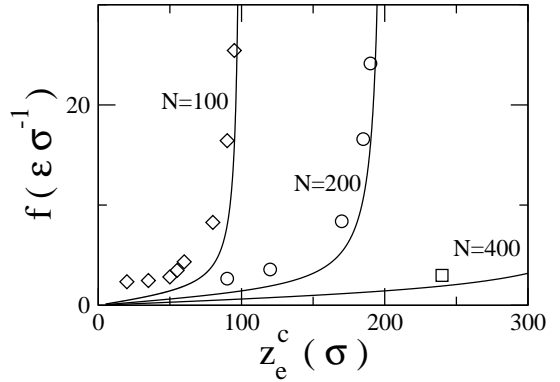}\\
     \caption{Force at the detachment point $z_e^c$ for different values of $w$ and
     $N$.
     Diamonds: $N=100$ $w=3,4,6,8,10,20,40,60$ from left to right.
      Circles: $N=200$ $w=4,8,10,20,40$.
     Square: $N=400$ $w=4$. Full lines are theoretical predictions.}
     \label{figure4}
\end{figure}
We finally investigate the globule structure in the regime of weak
adsorption in order to quantify its elastic deformation upon
stretching. We represent in Fig. 5 the monomer density  $\rho(z)$
and  energy $E(z)$ profiles  along  $z$  for different values of
$z_e$.  The density $\rho(z)$ presents  decaying density
oscillations similar to the one observed for a liquid  in the
presence of a solid substrate \cite{israelachvili92}.  The
presence of the density oscillations are due to the poor solvent
effect.We have checked that in the good solvent case there are no
density oscillations. As shown in Fig. 5 (top) the overall shape
of the density profile in the globule is invariant, up to a
normalization factor, as we stretch the chain. The layered
structure can therefore co-exist with the stretched chain. As the
monomers are extracted, the globule shape remains qualitatively
the same, independent of the size and of the tension as long as
the number of monomer in the globule is large enough. In Fig. 5
(bottom), we plot the energy density profiles $E(z)$ where $E(z)$
is the sum of the valence and of the Lennard-Jones interaction
energies. We can see from $E(z)$ plotted for different values of
$z_e$ that there exist three different regions. In the first
region ($z<3$) mostly corresponding to the first adsorbed layers,
the energy is independent of $z_e$. This is consistent with  the
above discussion on the negligible globule shape variation upon
stretching. A second region, roughly comprised between $z=3$ and
$z=8$, is formed between the globule and the chain. In this
matching region the top of the globule is elastically deformed and
pushed up by the tension. The overall vertical mechanical
equilibrium is ensured by the elevation  of the top of the
globule. The  small difference in the energy profiles computed for
$z_e=15$ and $z_e=20$ in the second region also demonstrates that
the elastic deformation is independent of the stretched chain
length. Finally the third region ($z>8$), corresponds to the
stretched chain with a very small  potential energy, confirming
our main theoretical hypothesis.\\
\begin{figure}
\includegraphics[width=0.35\textwidth]{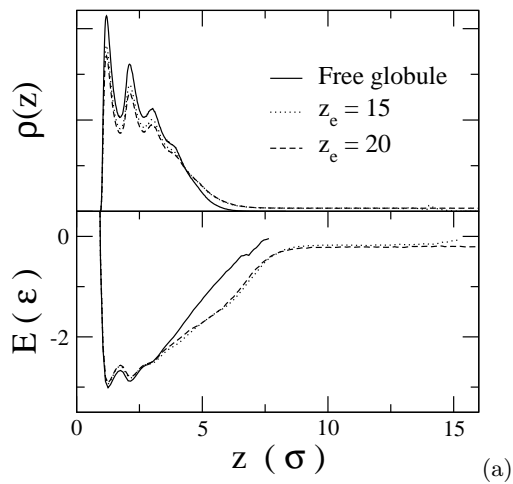}(a)\\
\caption{top:$N=100$, density profiles $\rho(z)$ showing layering
for a free globule and globules with $z_e=15$ and $20$ ($w=8$).
Bottom: energy profiles $E(z)$ for the same parameters.}
     \label{figure5}
\end{figure}
The picture we have presented above holds also for larger value of
$N$. Simulations which are not presented here have shown that
$f_p$ depends only weakly in $N$ and finally saturates. To
summarize, our simulations have shown that
there is a plateau  value for the force  that depends on the
substrate interaction $w$. In the strong adsorption limit the
un-stretched part of the polymer forms a 2D globule. A dynamical
coexistence between states with $N_c$  and $N_c+1$ monomers in the
chain has been observed. For weaker $w$ values the adsorbed
polymer forms a 3D globule wetting the substrate and elastically
deformed upon stretching. A simple theoretical model of the force
extension relation is presented and leads to a good agreement with
simulation results. The influence of the chain persistance length,
of the temperature and of  finite size effects will be presented
elsewhere. We hope that this work will motivate new experiments
 on the stretching of long molecules and bio-molecules. The weak
 adsorption regime will be investigated elsewhere.

\bibliography{cel}
\end{document}